\documentclass[amsfonts,amsmath,aps,prl,nofootinbib,reprint]{revtex4-2}
\usepackage[dvipdfmx]{graphicx}
\usepackage{bm,latexsym,amsmath,amssymb,amsfonts} 
\usepackage[breaklinks=true,colorlinks=true]{hyperref}
\usepackage{xcolor}
\usepackage{wasysym}
\usepackage{multirow}
\usepackage{enumitem}
\newcommand{\orcid}[1]{\href{https://orcid.org/#1}{\includegraphics[width=8pt]{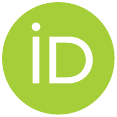}}}
\definecolor{royalblue}{rgb}{0.0, 0.0, 0.8}
\hypersetup{linkcolor=royalblue, citecolor=royalblue, urlcolor=royalblue}

\begin{document}

\def\Journal#1#2#3{\href{https://doi.org/#3}{{\it #1} #2}}
\def\arXiv#1#2{\href{https://arxiv.org/abs/#1}{arXiv:#1} [#2]}
        
\title{
Comment on ``Strong lensing and shadow of Ayon-Beato-Garcia (ABG)\\nonsingular black hole"}

\author{M.~Fahmi Fauzi\orcid{0009-0005-3380-6005}}
\email{muhammad.fahmi31@ui.ac.id}
\affiliation{Departemen Fisika, FMIPA, Universitas Indonesia, Depok 16424, Indonesia}

\begin{abstract}
    A recent article (Ramadhan \textit{et al.} Eur Phys J C 83:465, 2023) suggested that the Sagittarius A* black hole cannot be modeled by an Ay\'on-Beato-Garc\`ia regular black hole due to the incompatible shadow radius. However, they overlooked the distant-observer limit of the correction term in the effective geometry, which renders their calculation invalid. When evaluated correctly, the result reverses the conclusion of the paper. This comment highlights the flaw and demonstrates how to properly compute the shadow radius of such a black hole model using elementary geometry.
\end{abstract}

\maketitle

It was suggested by Ramadhan \textit{et al.}~\cite{Ramadhan:2023ogm} that the Sagittarius A* black hole (BH) cannot be modeled by an Ay\'on-Beato-Garc\`ia (ABG) regular black hole (RBH). Their calculation indicates that, when incorporating the photon effective geometry arising from nonlinear electrodynamics (NLED)~\cite{Novello2000}, the resulting shadow radius lies well below the $2\sigma$ and $1\sigma$ constraints from the Event Horizon Telescope combined with Keck and VLTI observational estimates obtained by Ref.~\cite{Vagnozzi:2022moj} (see also Ref.~\cite{EventHorizonTelescope:2022xqj}). The same conclusion was reached even for the uncharged case, which may appear reasonable given that the NLED Lagrangian does not reduce to Maxwell electrodynamics. However, an overlooked step in their shadow radius calculation renders the result invalid.

Let me briefly review the formalism and method used in Ref.~\cite{Ramadhan:2023ogm} to obtain the shadow radius of the BH. The seed BH metric is the Bardeen RBH~\cite{bardeen}, given by the line element
\begin{equation}
ds^2 = -f(r)dt^2 + f(r)^{-1}dr^2+h(r)r^2d\Omega^2,
\end{equation}
where
\begin{equation}
f(r) = 1-\frac{2Mr^2}{(r^2+q^2)^{\frac{3}{2}}}, \quad h(r)=1.
\end{equation}
Ay\'on-Beato and Garc\`ia~\cite{Ayon-Beato:2000mjt} showed that the Bardeen RBH solution is in fact a solution of the Einstein field equations when a NLED Lagrangian is incorporated. Moreover, it was proposed by Novello \textit{et al.}~\cite{Novello2000} that a NLED Lagrangian may introduce an effective geometry for photons, thereby modifying the metric tensor. For the NLED introduced in Ref.~\cite{Ayon-Beato:2000mjt}, this results in a modification of the two-sphere element, $h(r)$, given by
\begin{equation}
h_{ABG}(r)=\left[1-\frac{2(6q^2 - r^2)}{(q^2+r^2)}\right]^{-1}.
\label{eq. hr}
\end{equation}
One can see that for $q=0$, the effective geometry does not reduce to the Schwarzschild spacetime as $h_{ABG}(r)|_{q=0}=1/3$. Hereafter, I will refer to the effective geometry with $h(r)=h_{ABG}(r)$ as the \textit{ABGBH} spacetime.

The geodesic equation in Ref.~\cite{Ramadhan:2023ogm} is presented by Eq.~(14) of that paper. For convenience, however, let me instead use the form (see, e.g., Refs.~\cite{Fauzi:2024nta,Fauzi:2025rcc,Rohim:2025gxo})
\begin{equation}
\frac{d\phi}{dr}=\frac{1}{h(r)r^2}\left[b^{-2}-V(r)\right]^{-1/2},\quad
V(r)= \frac{f(r)}{h(r)r^2},
\label{eq. geodesic}
\end{equation}
where $V(r)$ is the photon effective potential and $b\equiv L/E$ is the impact parameter, with $L$ and $E$ being conserved quantities identified as the test particle’s angular momentum and energy, respectively. It can be readily verified that Eq.~(14) of Ref.~\cite{Ramadhan:2023ogm} is equivalent to Eq.~\eqref{eq. geodesic} used here.

\begin{figure}[htbp!]
	\centering
	\includegraphics[width=0.5\textwidth]{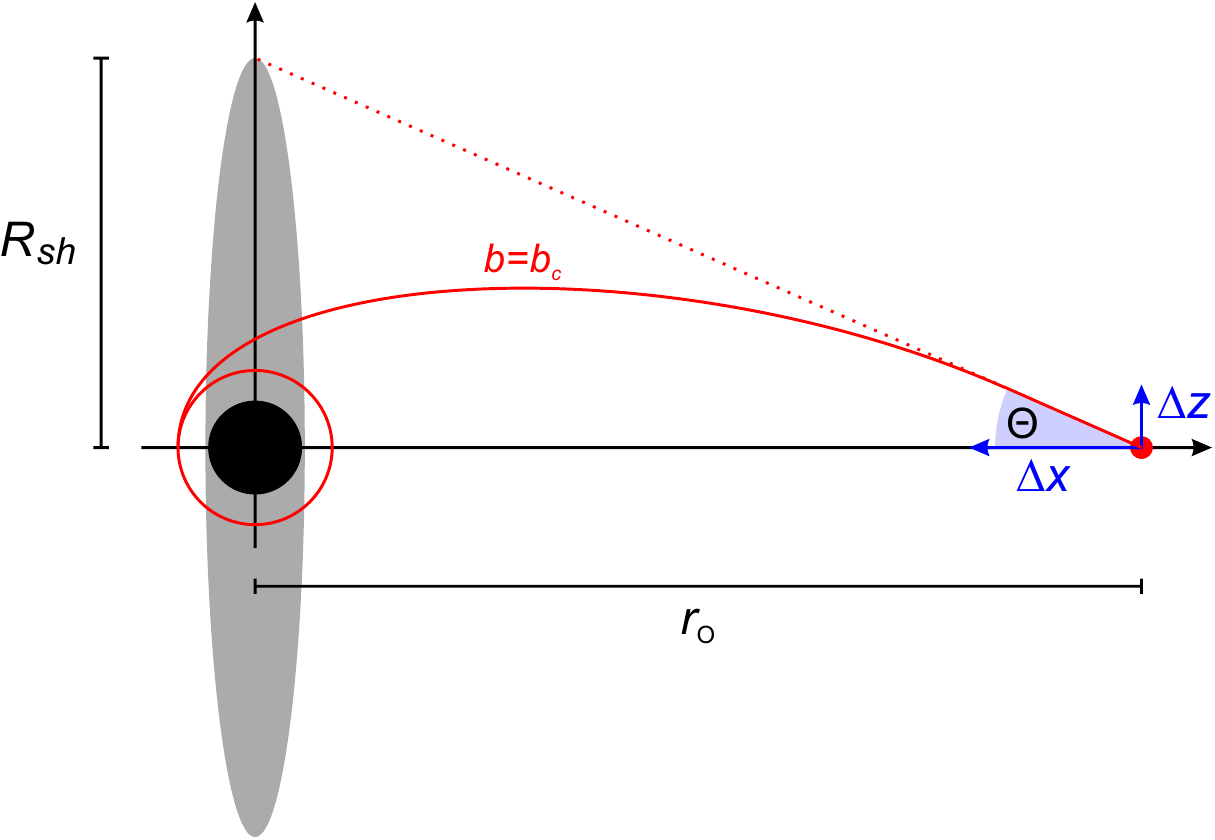}
	\centering
	\caption{The geometric setup for calculating the BH shadow radius, $R_{sh}$. The black filled circle represents the event horizon of the BH, and the red circle around it denotes the photon sphere. The red dot at $r_O$ marks the observer’s location.}
        \label{fig. geometric set shadow}
\end{figure}

A key quantity for computing the shadow radius is the critical impact parameter $b_c$, defined as the impact parameter of a photon orbiting the BH’s photon sphere. The radius of the photon sphere $r_{ps}$ can be determined by locating the extremum of the effective potential, \textit{i.e.} $\left.\partial_rV(r)\right|_{r=r_{ps}}=0$, and the critical impact parameter is then given by
\begin{equation}
b_c=\sqrt{\frac{1}{V(r_{ps})}}=r_{ps}\sqrt{\frac{h(r_{ps})}{f(r_{ps})}}.
\end{equation}
These relations are stated in Eqs.~(16) and (52) of Ref.~\cite{Ramadhan:2023ogm}.

Let me now present the basic concept of how the shadow radius can be computed using elementary geometry. In Fig.~\ref{fig. geometric set shadow}, I illustrate the schematic setup of the scenario. In short, the angle $\Theta$\textemdash identified as the \textit{angular radius} of the BH shadow \textemdash is given by (see, e.g., Refs.~\cite{Perlick:2021aok,Perlick:2018iye})
\begin{align}
\tan \Theta = &\frac{R_{sh}}{r_O}
=\lim_{\Delta x \to 0}\frac{\Delta z}{\Delta x} \equiv \left.\sqrt{\frac{h(r)r^2}{f(r)}}\frac{d\phi}{dr}\right|_{r=r_O,\;b=b_c}.
\end{align}
Substituting Eq.~\eqref{eq. geodesic}, one obtains
\begin{equation}
R_{sh}=r_Ob_c\sqrt{\frac{f(r_O)}{h(r_O)r_O^2-b_c^2f(r_O)}}.
\label{eq. rsh}
\end{equation}
In the limit of a distant observer ($r_O\to \infty$), if the spacetime is asymptotically flat, \textit{i.e.}, $\lim_{r\to \infty}f(r)=1$ and $\lim_{r\to \infty}h(r)=1$, Eq.~\eqref{eq. rsh} reduces to
\begin{equation}
R_{sh}=b_c,
\label{eq. rsh bc}
\end{equation}
which was used in Ref.~\cite{Ramadhan:2023ogm} to compute the shadow radius.

\begin{figure}[htbp!]
	\centering
	\includegraphics[width=0.4\textwidth]{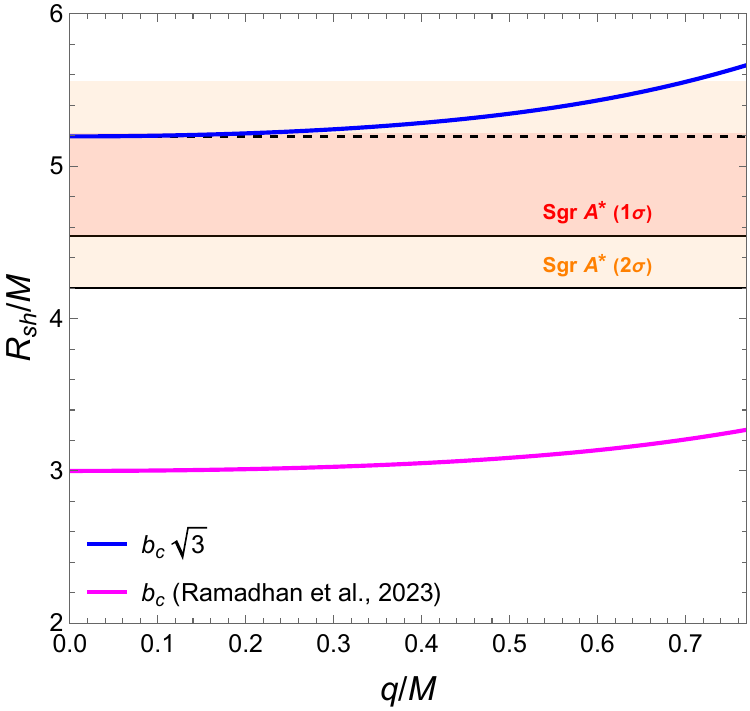}
	\centering
	\caption{The shadow radius obtained from Eq.~\eqref{eq. rsh bc sqrt3} (blue curve) and that presented in Ref.~\cite{Ramadhan:2023ogm} (magenta curve), shown together with the $1\sigma$ and $2\sigma$ constraints from the Sagittarius A* BH. The black dashed line represents the shadow radius of a Schwarzschild BH.}
        \label{fig. comparing shadow rad}
\end{figure}

While this relation has also been used in numerous studies (Ref.~\cite{Vagnozzi:2022moj} in particular), it is not valid within the ABGBH spacetime. The condition $\lim_{r\to \infty}h(r)=1$ does not hold for $h_{ABG}(r)$ in Eq.~\eqref{eq. hr}. In fact, one finds $\lim_{r\to\infty}h_{ABG}(r)=1/3$, and the shadow radius observed by a distant observer simply becomes
\begin{equation}
R_{sh}=b_c\sqrt{3},
\label{eq. rsh bc sqrt3}
\end{equation}
which introduces a factor of $\sqrt{3}$ that is missing in Eq.~\eqref{eq. rsh bc}. I compare the properly evaluated shadow radius with that obtained in Ref.~\cite{Ramadhan:2023ogm} in Fig.~\ref{fig. comparing shadow rad}. It can be directly seen that the shadow radius of the ABGBH lies within the Sagittarius A* $2\sigma$ constraint for $q/M\lesssim0.7$, and in the limit $q=0$, the Schwarzschild BH shadow radius is recovered.

Furthermore, it is plausible that the $\sqrt{3}$ scaling also extends to other observables, in particular the innermost image $\theta_\infty$ and the magnification $r_m$. One can see that for $q/M=0$, the value of $\theta_\infty$ and $r_m$ for the ABGBH reported in Table 1 of Ref.~\cite{Ramadhan:2023ogm} coincide with those of the Schwarzschild BH once rescaled by the $\sqrt{3}$ factor. However, a more complete treatment is required to validate this speculation.

In conclusion, the shadow radius\textemdash and possibly other strong lensing observables\textemdash of ABGBH reported in Ref.~\cite{Ramadhan:2023ogm} is inaccurate. A proper evaluation shows that the Sagittarius A* BH can indeed be consistent with the ABGBH model, thereby reversing the original claim.


\end{document}